\documentclass[useAMS,usenatbib,onecolumn]{mn2e}

\usepackage{amsmath}
\usepackage{epsfig}

\begin{document}

\title[On the paucity of Fast Radio Bursts at low Galactic latitudes]{On the paucity of Fast Radio Bursts at low Galactic latitudes}

\author[Jean-Pierre Macquart and Simon Johnston]{Jean-Pierre Macquart$^{1,2}$\thanks{E-mail:
J.Macquart@curtin.edu.au} and Simon Johnston$^{3}$ \\
$^{1}$ICRAR/Curtin University, Curtin Institute of Radio Astronomy, Perth WA 6845, Australia\\
$^{2}$ARC Centre of Excellence for All-Sky Astrophysics (CAASTRO) \\
$^{3}$CSIRO Astronomy and Space Science, P.O. Box 76, Epping, NSW 1710, Australia}


\pagerange{\pageref{firstpage}--\pageref{lastpage}} \pubyear{2015}

\maketitle

\label{firstpage}

\begin{abstract}
We examine the effect of Galactic diffractive interstellar scintillation as a means of explaining the reported deficit of Fast Radio Burst (FRB) detections at low Galactic latitude.   We model the unknown underlying FRB flux density distribution as a power law with a rate scaling as $S_\nu^{-5/2+\delta}$ and account for the fact that the FRBs are detected at unknown positions within the telescope beam.  We find that the event rate of FRBs located off the Galactic plane may be enhanced by a factor $\sim 30-300$\% relative to objects near the Galactic plane without necessarily affecting the slope of the distribution.  For FRBs whose flux densities are subject to relatively weak diffractive scintillation, as is typical for events detected at high Galactic latitudes, we demonstrate that an effect associated with Eddington bias is responsible for the enhancement.  The magnitude of the enhancement increases with the steepness of the underlying flux density distribution, so that existing limits on the disparity in event rates between high and low Galactic latitudes suggest that the FRB population has a steep differential flux density distribution, scaling as $S_\nu^{-3.5}$ or steeper.  Existing estimates of the event rate in the flux density range probed by the High Time Resolution Universe (HTRU) survey overestimate the true rate by a factor of $\sim 3$. 
\end{abstract}


\section{Introduction}
Fast Radio Bursts (FRBs) are a new class of millisecond-timescale, decimeter-wavelength radio transients that potentially occur at cosmological distances (Lorimer et al.\,2007; Thornton et al.\,2013; Spitler et al.\,2014).  The dispersion measures of FRBs exceed the expected contribution from the Galactic Interstellar Medium (ISM) by factors of several, leading to the suggestion that the Inter-Galactic Medium (IGM) contributes the bulk of the dispersion measure.  Comparison against IGM models (Ioka 2003; Inoue 2004; see also McQuinn 2013) implies that these bursts are detected at redshifts up to $z \approx 1$.  Moreover, at least two of the bursts exhibit clear evidence of temporal smearing caused by electron density inhomogeneities that must be external to the Galaxy; the $\sim 1\,$ms timescale of the observed smearing exceeds the expected Galactic contribution by at least two orders of magnitude (Cordes \& Lazio 2002 (NE2001)).  

A puzzling characteristic of the FRB population is that the detection rate increases with Galactic latitude. This result is based on the dearth of FRB detections in the low- and mid-latitude components of the 1.4\,GHz Parkes High-Time Resolution Universe (HTRU) survey (Keith et al. 2010), despite the large fraction of survey time spent at these latitudes. Petroff et al. (2014) conclude that, even after taking into account selection effects such as enhanced sky temperature, increased dispersion measure smearing and excess scattering at low Galactic latitudes, the FRB distribution is non-isotropic with 99\% confidence (see also Burke-Spolaor \& Bannister 2014).

It has been suggested that interstellar scattering offers an explanation for this latitude dependence (Petroff et al.\,2014).  The specific suggestion is that diffractive interstellar scintillation amplifies the emission from some FRBs so that events which would ordinarily be too weak to be detected are pushed above the threshold of detectability.  Similar suggestions have been advanced in similar contexts, notably in relation to the effect of diffractive scintillation on SETI signals (Cordes, Lazio \& Sagan 1997) and the effect of gravitational lensing on quasar number counts (Turner 1980).

There are several {\it prima facie} motivations to favour this suggestion.
\begin{itemize}
\item[1] The implied distribution appears unphysical if it is intrinsic to
the progenitor population. Extragalactic sources should exhibit no
latitude dependence whereas Galactic sources confined to the disk
should exhibit an excess of events at low latitudes. One can imagine
a Galactic population consisting of a very local population or
conversely a population in an extended halo but even these populations
show some biases towards low Galactic latitudes (see, e.g., the debate
over the progenitor population of gamma-ray bursts in the 1990s as summarised by Fishman \& Meegan (1995)).
\item[2]  No known absorption effect can quench the radiation in such a latitude-dependent manner at the frequencies at which FRB radiation is detected.  Free-free absorption can in principle influence the properties of the observed population, but this is known not to be an important consideration for other types of Galactic populations at this frequency (e.g. pulsars).
\item[3] Insterstellar scintillation qualitatively explains the observed Galactic latitude dependence.  The decorrelation bandwidth of the scintillation at high Galactic latitudes approximately matches the value required for scintillation to amplify the radiation; specifically it can be comparable to or larger than the observing bandwidth of the HTRU survey.  Large amplifications due to scintillation are only possible under this circumstance.  For the stronger scintillations closer to the Galactic plane where the decorrelation bandwidth is considerably smaller than the observing bandwidth, the average over a large number of diffractive scintillations across the observing band would result in no nett amplification of the signal.  Thus, at low Galactic latitudes the observed flux densities would approach the intrinsic flux densities of the events themselves, which may not exceed the threshold for detectability with HTRU.
\end{itemize}

In this paper we examine the scintillation hypothesis critically.  A particular implication is that the scintillation should alter the observed FRB flux density distribution in a specific manner.  We derive the observed flux density distribution in terms of the intrinsic flux density distribution and the probability distribution of the amplification provided by diffractive scintillation.  A quantitative model is introduced in \S\ref{sec:ScintEnhance}.  The  discussion is confined  to detections made at the Parkes radiotelescope, which account for all but one of the FRBs detected to date, since it is not yet possible to ascertain whether detections at other telescopes exhibit a similar Galactic latitude dependence.
In \S\ref{sec:Compare} we assess this model as an explanation of the FRB Galactic latitude dependence and examine its implications for FRB surveys.  Our conclusions are presented in \S\ref{sec:Conclusions}.

\section{Scintillation enhancement of FRB flux densities} \label{sec:ScintEnhance}

We consider the amplification of radiation due to diffractive scintillation caused by the turbulent ISM of our Galaxy as a possible explanation of the observed latitude dependence of FRB detections.  
For diffractive scintillation to viably enhance the radiation observed from an FRB two conditions must be satisfied:  

First, the decorrelation bandwidth of the interstellar diffractive scintillations must be comparable to or larger than about half the observing bandwidth.  The random amplifications due to diffractive scintillation are correlated only over a finite decorrelation bandwidth, $\Delta \nu_{\rm dc}$.  If the observing bandwidth, $\Delta \nu_b$, spans a large number of decorrelation bandwidths, no nett enhancement would be observed.  At observing frequencies $1.2-1.5\,$GHz where FRB detections have been made, large decorrelation bandwidths are typically only encountered at high Galactic latitudes.  For instance, the NE2001 (Cordes \& Lazio 2002) scattering model predicts that the decorrelation bandwidth is in the range $\Delta \nu_{\rm dc} = 1.5-3.5\,$MHz at $|b| \approx 30^\circ$ at 1.5\,GHz; this is much smaller than the $\Delta \nu_b = 320\,$MHz used in the FRB detections made with Parkes multibeam receiver.  

The NE2001 model is poorly constrained at high Galactic latitudes and 
there is substantial evidence that regions of weaker scattering, with commensurately larger decorrelation bandwidths, do in fact exist at high Galactic latitudes.  This is derived, for instance, from the scattering properties of intra-day variable quasars viewed through lines of sight $|b|>30^\circ$ off the Galactic plane.   The $\approx 3$-$5$\,GHz transition frequency between weak and strong scintillation deduced for many well-studied intra-day variable quasars (Dennett-Thorpe \& de Bruyn 2003, Macquart et al.\,2000; Kedziora-Chudczer et al.\,1997; Bignall et al.\,2003) implies decorrelation bandwidths of $40$-$190$\,MHz at 1.5\,GHz.  There are several high latitude pulsars with distances $\ga 300\,$pc (i.e.\,above most of the turbulent electron layer) with values of $\Delta \nu_{\rm dc}$ at 1.4\,GHz which exceed $\approx 300\,$MHz, notably B1237$+$25 and B0031-07 (Bhat, Gupta \& Rao 1998).
Moreover, the absence of discernible frequency structure in the observed spectra of the brightest reported FRBs (Thornton et al. 2013) itself provides evidence that the decorrelation bandwidth must either be smaller than the channel resolution (390\,kHz), or must be larger than the 320\,MHz HTRU observing bandwidth for the lines of sight probed by the FRBs detected to date.   \\

The second condition is that scattering by turbulence external to our Galaxy must not corrupt the properties of the incident radiation in a manner that destroys the ability of the ISM to amplify the signal.  This effect is relevant because it appears that at least two FRBs are subject to significant scattering beyond our Galaxy: the $\tau \sim 1\,$ms pulse smearing observed in both the Lorimer burst and FRB\,110220 is $\sim 10^4$ times larger than can be plausibly accounted for by scattering within the Galaxy (Macquart \& Koay 2013; Luan \& Goldreich 2014).  

The essential requirement is that the extragalactic scattering does not alter the coherence properties of the radiation in a manner that prevents interstellar scattering from modulating the incident signal.  In practice, this means that the spatial coherence length of the wavefront scattered by extragalactic turbulence, $s_{\rm IGM}$, is no smaller than the size of the scattering disk associated with the interstellar scintillations, $s_{\rm scat}$.  For $\Delta \nu_{\rm dc}$ of $300\, \Delta_{300} \,$MHz at 1.4\,GHz and an effective distance to the interstellar turbulence of $D_{\rm kpc}\,$kpc, the size of the scattering disk is $s_{\rm scat} = 2 \times 10^9 \,\Delta_{300}^{-1/2} D_{\rm kpc}^{1/2}\,$m.

We estimate $s_{\rm IGM}$ using eqs.(12) \& (14) of Macquart \& Koay (2013) at 1.4\,GHz to be
\begin{eqnarray}
s_{\rm IGM} = \frac{\lambda}{2 \pi \sqrt{c \, \tau (1+z_L)} } 
\left( \frac{D_{\rm L} D_{\rm S}}{ D_{\rm LS} } \right)^{1/2} \sim \left( \frac{\tau}{1\,{\rm ms}} \right)^{-1/2}
\begin{cases} 
3 \times 10^8 \left( \frac{ D_S}{1\,{\rm Gpc}} \right)^{1/2}\,{\rm m}, & D_{\rm LS} \approx D_{\rm L}, \\
3 \times 10^{11} \left( \frac{D_L}{1\,{\rm Gpc}} \right)  \left( \frac{ D_{\rm LS}}{1\,{\rm kpc}} \right)^{-1/2}\,{\rm m}, &  D_{\rm LS} \ll D_{\rm S}, \\
\end{cases}
\end{eqnarray}
where the scattering occurs an angular diameter distance $D_{\rm L}$ (with associated redshift $z_L$) from the observer due to an event at a distance $D_{\rm S}$, and the angular diameter distance  between the source and scattering plane is $D_{LS}$.

Since $s_{\rm IGM} \ga s_{\rm scat}$ we see that in most cases the effect of extragalactic scattering is insufficient to corrupt the radiation incident upon the interstellar medium.  This condition is easily met if the scattering is associated with the host galaxy (i.e. $D_{\rm LS} \ll D_{\rm S}$).  The one possible exception is if the scattering time substantially exceeds a millisecond and if the extragalactic scattering takes place at distances intermediate to the FRB itself.

We also note in passing that there is a decorrelation bandwidth associated with the extragalactic scattering.  However, its value of $\sim 1\,$kHz for scattering times $\sim 1\,$ms is much smaller than the spectral resolution of HTRU filterbank, rendering these spectral variations unobservable.

In summary, we conclude that both conditions are plausibly satisfied for most of the FRB population detected to date, and it is therefore possible in principle that interstellar diffractive scintillation has a substantial effect on the observed flux densities of detected FRBs.



\subsection{Flux density distribution}

To determine the flux density distribution of events incident upon the telescope, we must model both the effect of the scintillations and the initial (unperturbed) flux density distribution of the transient events.

In the regime of strong diffractive scintillation, the probability of observing an amplification, $a$, over the mean source flux density from a single scintle is (Mercier 1962; Salpeter 1967, see also Narayan \& Goodman 1989; Gwinn et al. 1998)
\begin{eqnarray}
p_{a,1}(a) =  \exp \left( - a \right). \label{pa1}
\end{eqnarray}
In the opposite limit in which a large number of scintles, $N$, contribute to the flux density across a finite observing bandwidth the central limit theorem implies the amplification distribution approaches a normal distribution,
\begin{eqnarray}
p_{a,N}(a) = \frac{1}{\sqrt{2 \pi \sigma_N^2}} \exp \left( - \frac{(a-1)^2}{2 \sigma_N^2}\right), \label{paN}
\end{eqnarray}
with a mean of unity and a standard deviation $\sigma_N \sim 1/\sqrt{N}$.

The distribution of observed event flux densities subject to enhancement by scintillation is computed by examining the probability density of the variate $Z = S_\nu \times a$, the product of the amplification with $S_\nu$, the intrinsic burst flux density.  For a differential distribution of event flux densities, $p_S (S_\nu)$, between some minimum and maximum flux densities $S_{\rm min}$ and $S_{\rm max}$ respectively, the observed distribution is
\begin{eqnarray}
p_Z (Z) &=& \int_{S_{\rm min}}^{S_{\rm max}}  p_S(S_\nu) p_{a,1} \left( \frac{Z}{S_\nu} \right) \frac{dS_\nu}{S_\nu} \label{pZeq}.  \label{VariateMultiply} 
\end{eqnarray}

To understand the implications of this basic result, we parameterize the differential flux density distribution in terms of a power law,
\begin{eqnarray}
p_S(S_{\nu}) = \begin{cases}
K S_{\nu}^{-5/2 + \delta} & S_{\rm min} < S_{\nu} < S_{\rm max}, \\
0 & \mbox{otherwise}, \\
\end{cases} \qquad K = \frac{r \left(\delta -\frac{3}{2} \right)}{S_{\rm min}^{-{3/2}+\delta} - S_{\rm max}^{-3/2+\delta}}, \label{pS}
\end{eqnarray}
where $r$ is the overall detection rate integrated over the entire flux range over which events occur, $S_{\rm min} < S_{\nu} < S_{\rm max}$.  

\begin{figure}
\centerline{\epsfig{file=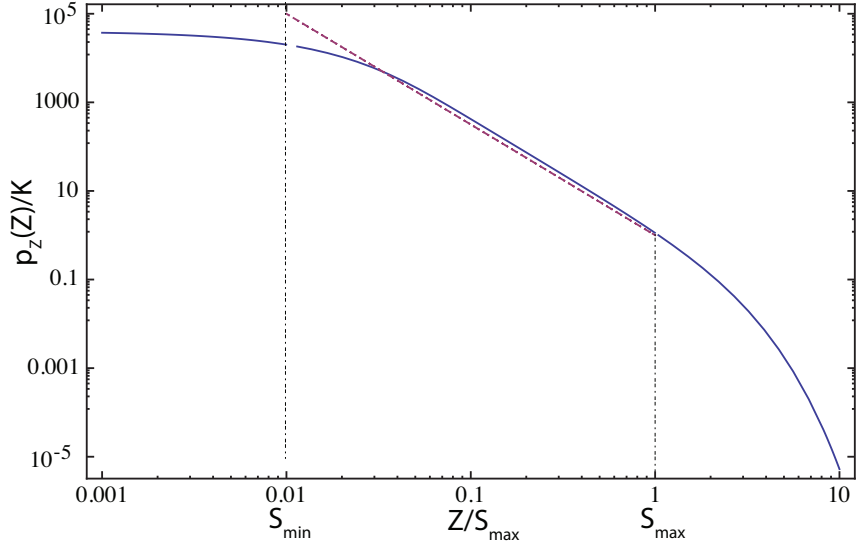, scale=0.7}}
\caption{The distribution of observed flux densities $p_Z(Z)$ (blue solid line) for an initial flux density distribution (purple dashed line) that is nonzero over the range $S_{\rm min} < Z < S_{\rm max}$ and with $\delta =0$.  The effect of the diffractive scintillations is to draw out the high end of the distribution into a tail that decreases like $Z^{-1} \exp(-Z)$, increase the differential event counts over the range $S_{\rm min} \ll Z \la S_{\rm max}$ and to extend the low luminosity component of the distribution to zero flux density. } \label{fig:pZ}
\end{figure}

A finite flux density cutoff at $S_{\rm max}$ is introduced as a means of investigating the effect of scintillation enhancement if a steep decline is evident at high flux densities.  This is introduced as a means of investigating an aspect of the specific suggestion made in Petroff et al.\,(2014).  However, we remark that $S_{\rm max}$ is a free parameter, and may be taken to be arbitrarily large.  Indeed, the physical origin of such a cutoff is unclear: a more physical parameterisation would consider events distributed between lower and upper cutoff intrinsic luminosities in conjunction with some distribution of event distances.  Even for an extreme luminosity distribution in which FRBs were standard candles, the distribution would exhibit no sharp cutoff if the events were homogeneously distributed.  
In general, the value of $S_{\rm max}$ is finite if FRBs occur only at cosmological distances (i.e. at a certain minimum distance) and there is a finite maximum burst luminosity.  The distribution scales as $S_{\nu}^{-5/2}$ if the progenitor population does not evolve as a function of redshift and effects due to the curvature of spacetime across cosmological distances are neglected.  The factor $\delta$ takes into account evolution in the progenitor population and the non-Euclidean geometry of the Universe at $z \ga 1$: similar effects in the quasar population give rise to number counts with $-0.5 \la \delta \la 0.5$ (see, e.g., Wall 1980; Wall 1994).  
Evaluating eq.(\ref{VariateMultiply}) for enhancement by a single scintle, we obtain
\begin{eqnarray}
p_{Z,1}(Z) &=& K Z^{-5/2 + \delta} \left[ \Gamma_{5/2 -\delta} \left( \frac{Z}{S_{\rm max}} \right)
- \Gamma_{5/2 -\delta} \left( \frac{Z}{S_{\rm min}} \right) \right]
, \label{FullS0soln} \\
&\approx& K
\begin{cases}
S_{\rm min}^{-5/2 + \delta}/(\frac{5}{2} - \delta)  & Z \ll S_{\rm min}, \\
Z^{-\frac{5}{2} +\delta} \Gamma \left( \frac{5}{2} - \delta \right)   	& S_{\rm min} \ll Z \la S_{\rm max}, \\
 S_{\rm max}^{-\frac{5}{2}+\delta} \left( \frac{Z}{S_{\rm max}} \right)^{-1} e^{\frac{-Z}{S_{\rm max}} }  &
Z \gg S_{\rm max},  \\
\end{cases}
\end{eqnarray}
where $\Gamma_{a}(Z) = \int_Z^\infty t^{a-1} e^{-t} dt$ is the incomplete gamma function.  The effect of scintillation is to enhance the event rate for all flux densities $Z \gg S_{\rm min}$, as illustrated in Figure \ref{fig:pZ}.  It introduces three significant effects: (i) it extends the distribution beyond $S_{\rm max}$ into an exponentially-decreasing tail, thus pushing a small fraction of events near $S_{\rm max}$ to yet higher flux densities, (ii) it enhances the event rate over the range $S_{\rm min} \ll Z \la S_{\rm max}$ by a factor $\Gamma(5/2+\delta)$ but the distribution retains the same power-law index as the original distribution, and (iii) it draws a portion of the low flux density distribution near $S_{\rm min}$ into a flat region that extends down to zero.  It is straightforward to verify that, as expected, the event rate integrated over the distribution remains identical to the original rate.  

The distribution extensions (i) and (iii) are, respectively, attributable to the fact that scintillation amplifies the flux density of a small fraction of events near $S_{\rm max}$ and that it similarly de-amplifies a fraction of events near $S_{\rm min}$.   The enhancement associated with (ii) is more subtle, and may be regarded as the effect of Eddington bias: because the distribution decreases steeply with flux density any effect that scatters objects in flux density preferentially scatters more objects from low to high flux density than vice versa.  It follows that the greater the steepness of the distribution (i.e. the greater the value of $-\delta$), the greater the enhancement in event rate, $\Gamma(5/2-\delta)$, due to this bias.



For completeness, we also consider the flux density distribution where a large number of scintles, $N$, contribute to the overall measurement of the flux density across the observing bandwidth, using eq.(\ref{paN}):
\begin{eqnarray}
p_{Z,N}(Z) = K Z^{-\frac{5}{2}+\delta} \left\{ \frac{2^{-\frac{1}{4} - \frac{\delta}{2}}}{\sqrt{ \pi}}    \sigma_N^{\frac{1}{2} - \delta} 
\left[ 
\sigma_N  \Gamma \left(\frac{5 - 2 \delta }{4}\right) \,
   _1F_1\left(\frac{2 \delta -3}{4};\frac{1}{2};\frac{-1}{2
   \sigma_N ^2}\right)+\sqrt{2} \Gamma \left(\frac{7- 2 \delta }{4}\right) \, _1F_1\left(\frac{2 \delta
   -1}{4};\frac{3}{2};\frac{-1}{2 \sigma_N ^2}\right)
\right] \right\},
\end{eqnarray}
where $_1 F_1$ is a confluent hypergeometric function, and we have taken the limits $S_{\rm min}=0$ and $S_{\rm max} \rightarrow \infty$ in order to make the problem analytically tractable.  The expression in the curly brackets represents the correction to the event rate over the non-scintillating signal.

\begin{figure}
\centerline{\epsfig{file=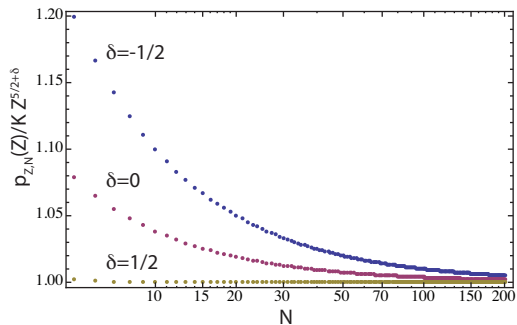, scale=0.5}}
\caption{The amplification of the event rate with the number of scintles that contributes to the flux density in the limit in which $N$ is large and $S_{\rm min} \rightarrow 0$ and $S_{\rm max} \rightarrow \infty$.} \label{fig:approachLim}
\end{figure}

In the limit $N \rightarrow \infty$ this distribution approaches the intrinsic distribution given by eq.(\ref{pS}).  For finite values of $N$, there is still some small increase in the event rate over the range $S_{\rm min} \ll Z \la S_{\rm max}$, but this diminishes rapidly with $N$, as shown in Figure \ref{fig:approachLim}.  The important point is that the overall enhancement is less than $\approx 5$\% for the small scintillation bandwidths typical of diffractive scintillation closer than $30^\circ$ to the Galactic plane where one has $N \ga 30$ for typical HTRU observing parameters.

\subsection{Flux density distribution measured with a single-pixel or multibeam receiver}

A complication arises in comparing the observed flux density distribution of FRBs with foregoing results because all FRB detections have hitherto been made with non-interferometric radio telescopes. Each event is detected at an unknown angular distance from the beam centre, and the observed signal-to-noise is attenuated by an unknown amount according to the beam shape.  However, although the effect of beam attenuation is unknown for any given event, it is nonetheless possible to derive its statistical effect on a population of events.  We quantify this effect here.  Our analysis is informed by two aspects of the HTRU survey: 
\begin{itemize}
\item[1]  With the exception of the Lorimer burst (Lorimer et al. 2007), each FRB was detected in only a single beam of the Parkes multibeam receiver, with no detections in adjacent beams at a significance exceeding $\approx 5 \sigma$.  The beam centres of the Parkes multibeam receiver are placed two full-width-at-half-maximum (FWHM) beamwidths apart, hence we deduce that every detected FRB occurred at an angular separation no greater than this distance from the beam centre. This argument is not rigorously true for events detected by one of the outer mulitbeam receivers because the event may have occurred on the outerward edge of the beam.   However, the number of FRBs detected in the outer beams of the Parkes multibeam receiver compared to those found in the inner beams is inconsistent with detections far beyond the primary beam pattern.\\
\item[2] The beam shape is frequency dependent, and if the FRB occurred beyond the first null of the beam at any frequency within the 1180-1500\,MHz observing range, the spectrum would exhibit extremely large gradients across the band and may even reverse slope.  The absence of such spectral features in Parkes FRBs reported to date suggests that no event was detected at Parkes at an angular separation close to the first beam null or beyond it.  (The FRB detected by Spitler et al.\,(2014) at Arecibo demonstrates the large spectral gradients possible when the event is detected far from the beam centre.)
\end{itemize}

We would like to compute the probability distribution of the attenuation.  We consider the attenuation associated with events whose locations fall within the half power point of a gaussian beam,
$B_{\rm gauss}(\theta ) = \exp ( - \theta^2/2 \theta_b^2 )$,  
where $\theta_b$ is a measure of the beamwidth.  The beamshape of every one of the 13 Parkes beams is distinct, because each off-axis beam suffers from effects of asymmetry caused by coma and diffraction of the radiation around the telescope feed legs.  However, the effects of beam asymmetry are mitigated substantially by the fact that we are only interested in events that occur well within the first beam null.  
Given the large uncertainties associated with the measured FRB flux density distribution at present, we are justified in deriving the attenuation probability distribution only to first order assuming this approximation. 

The probability that an event occurs a distance $\theta$ from the beam centre is proportional to 
$2 \pi \theta d\theta$, and the normalised probability distribution for an event that occurs out to a maximum distance $\theta_{\rm edge}$ is $p_\theta (\theta) d\theta= 2 \theta d\theta / \theta_{\rm edge}^2$. The probability of detecting an event with beam attenuation factor, $y$, is the probability of obtaining that value of $\theta$ which corresponds to $y=B (\theta)$.  We derive the distribution of beam attenuations by changing the variable of the probability distribution $p_\theta(\theta)$ to $B(\theta)$, and the attenuation probability density $P_y(y)$ is derived from $p_\theta$ using the relation,
\begin{eqnarray}
P_y(y) = \frac{p_\theta [g(y)] }{\left| B' [g(y)] \right|},
\end{eqnarray}
where $g(y) = \theta$ is the inverse function of $B(\theta)$.  Considering that event detections are confined to within two half-beamwidths from the pointing centre, namely $\theta_{\rm edge} = 2 \theta_b \sqrt{2 \ln 2}$, the resulting attenuation distribution is, 
\begin{eqnarray}
P(y)  = \frac{1}{(4 \ln 2)\, y}, \qquad \frac{1}{16} <y< 1. 
\end{eqnarray}
The cutoff at $y=1/16$ corresponds to the fraction of power received two half-beamwidths from the pointing centre.  The average attenuation is $\langle y \rangle = 0.39$.

\begin{figure}
\centerline{\epsfig{file=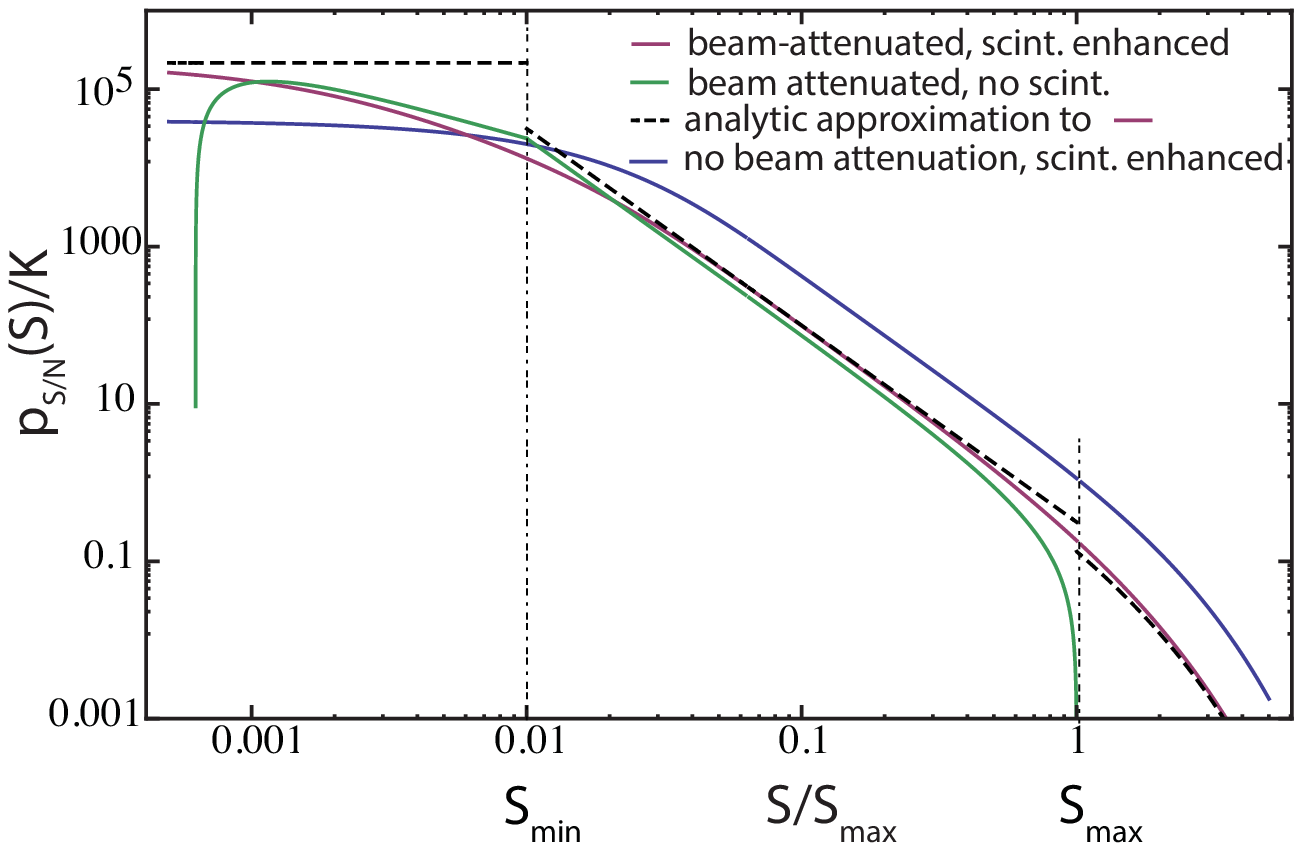,scale=0.7}
\epsfig{file=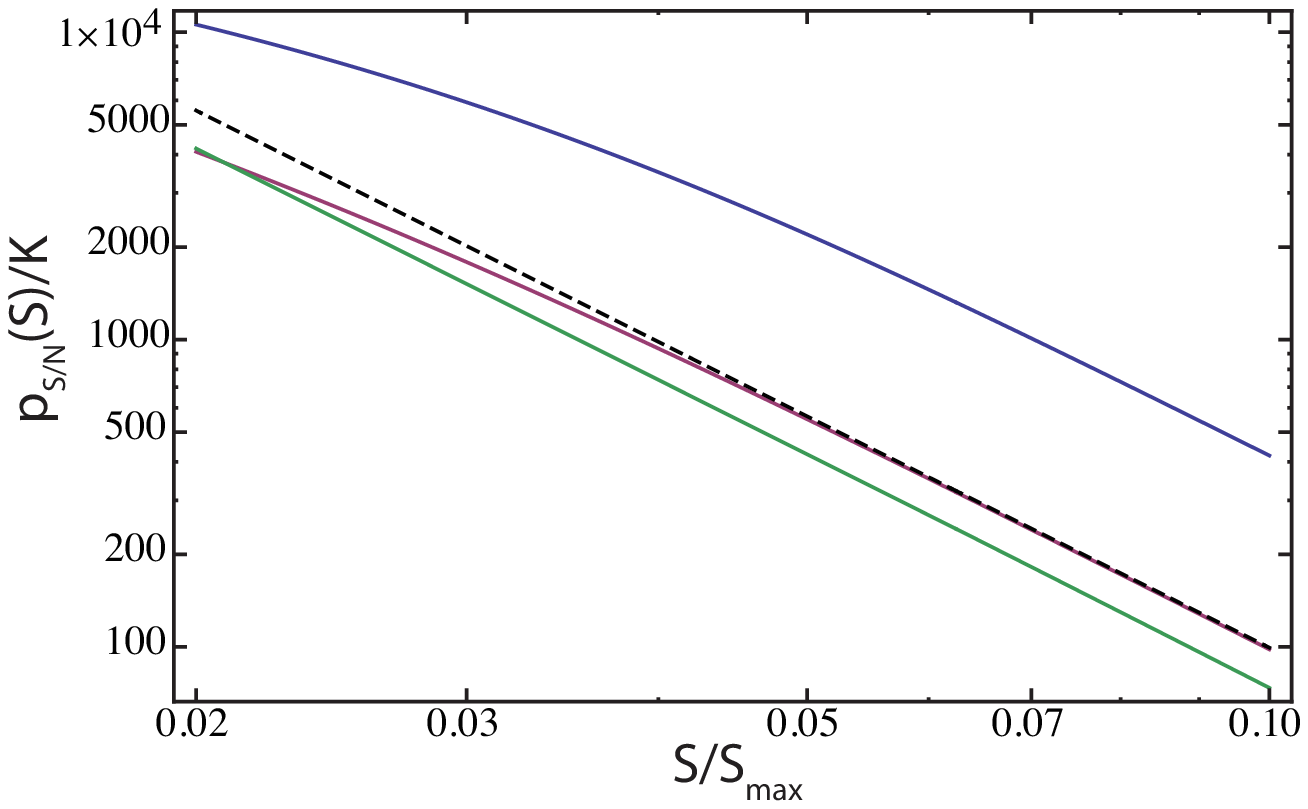,scale=0.7}}
\caption{A graphic illustration of the effect of beam attenuation on the distribution of observed event flux densities.  The distribution $p_{S/N}({\cal S})$ (purple) is the distribution measured at the telescope for an intrinsic distribution $p_{Z,1}(Z)$ (given by eq.\,(\ref{FullS0soln}), blue curve), taking into account the fact that events may be detected anywhere within the half-power point of the beam.  The calculation assumes a Gaussian beam shape.  The analytic approximations to $p_{S/N}({\cal S})$ in the limits ${\cal S} \ll S_{\rm min}$, $S_{\rm meas} \ll {\cal S} \la  S_{\rm max}$ and ${\cal S} \ga S_{\rm max}$, given by eq.\,(\ref{SmeasApprox}), are shown by the dashed line.   The beam-attenuated flux density distribution follows the same slope as the underlying distribution over the range $S_{\rm min} \ll {\cal S} \la S_{\rm max}$ but is offset from it.  The green curve shows the flux density distribution for events whose flux densities are not affected by scintillation.  The right plot displays a zoomed region of the left panel to demonstrate the difference in the amplitudes of the various distributions.} \label{fig:AttenDistn}
\end{figure}

The equivalent flux-density, ${\cal S}$, at which any given event is detected is proportional to the product of two random variates: the incident flux density (after amplification by scintillation) and the telescope attenuation, $y$.  For a population whose flux densities are initially distributed according to a power law, altered by diffractive scintillation (see eq.(\ref{FullS0soln})), and then subject to the random effects of beam attenuation, the measured equivalent flux density distribution is, using eq.(\ref{VariateMultiply}),
\begin{eqnarray}
p_{\rm S/N} \left( {\cal S} \right) &=&  \frac{1}{4 \ln 2} \int_{1/16}^1 
	p_{Z,1}  \left( \frac{{\cal S}}{y}  \right) \frac{dy}{y^2} \\
	&=& \frac{K {\cal S}^{-5/2 + \delta}}{ \left( \frac{3}{2} - \delta \right) 4 \ln 2}  \left[ G({\cal S},S_{\rm max}) -  G({\cal S} ,S_{\rm min}) \right], \\
	&\null& \qquad \mbox{where  } G({\cal S},S_0)  =  \Gamma_{\frac{5}{2} -\delta} \left( \frac{\cal S}{S_0} \right) 
	- 2^{4 \delta-6} \Gamma_{\frac{5}{2} -\delta} \left( \frac{16 \,{\cal S}}{S_0} \right) +
	\left( \frac{{\cal S}}{S_0}\right)^{\frac{3}{2}-\delta} \left( e^{-\frac{16 {\cal S}}{S_0}} - e^{-\frac{{\cal S}}{S_0}} \right). 
\end{eqnarray} 
%
%
%
The behaviour of this distribution is approximated as follows:
\begin{eqnarray}
p_{S/N} \left( {\cal S} \right) \approx    \frac{K \, {\cal S}^{-5/2+\delta}}{4 \ln 2}  
\begin{cases}
\frac{ 15 }{5/2-\delta} {\cal S}^{5/2-\delta} \left[ S_{\rm min}^{-5/2+\delta} - S_{\rm max}^{-5/2+\delta} \right]   &  {\cal S} \ll S_{\rm min}, \\
\Gamma \left( \frac{3}{2} - \delta \right) \left[ 1 - 16^{-3/2 + \delta} \right]  & S_{\rm min} \ll {\cal S} \la S_{\rm max}, \\ 
\left( \frac{\cal S}{S_{\rm max}} \right)^{1/2-\delta} \left[ e^{-{\cal S}/S_{\rm max}} - \frac{1}{16} e^{-16 {\cal S}/S_{\rm max}} \right]  & {\cal S} \ga S_{\rm max}. \\ 
\end{cases}
\label{SmeasApprox}
\end{eqnarray}
A plot of the distribution of the equivalent flux densities, $p_{S/N} ({\cal S})$, is shown in Figure \ref{fig:AttenDistn} along with the intrinsic (scintillation-affected) flux density distribution $p_{Z,1}(Z)$ and the analytic approximation to the distribution described by eq.(\ref{SmeasApprox}).  

In the regime $S_{\rm min} \ll {\cal S} \la S_{\rm max}$ the slope of the distribution remains the same as the underlying distribution.  However, in the regime ${\cal S} \gg S_{\rm min}$ the distribution of equivalent flux densities is systematically lower than the original distribution, $p_{Z,1}$, as a result of the fact that each event detected is subject to some degree of attenuation.  In essence, this is because no event occurs precisely at the beam centre.  For a given attenuation $1/16 < y <1$, the number of events detected at a given flux density ${\cal S}$ is the number of events whose flux density incident on the telescope is actually ${\cal S}/y$.   

For comparison with the scintillation-enhanced distribution, we derive the equivalent flux density distribution for events that are not subject to appreciable enhancement by diffractive scintillation.  As noted above, this situation applies to events located within $\sim 30^\circ$ of the Galactic plane, where a large number of scintles contribute to the observed flux density, since the decorrelation bandwidth of the diffractive scintillations is much smaller than the observing bandwidth of the HTRU survey.  The S/N distribution is 
\begin{eqnarray}
p_{S/N}({\cal S}) &=& \frac{1}{4 \ln 2} \int_{1/16}^1 p_S \left( \frac{\cal S}{y}\right) \frac{dy}{y^2} \\
&=&  \frac{K \, {\cal S}^{-5/2+\delta} }{4 \ln 2 \left(\frac{3}{2} - \delta \right)} \begin{cases}
\left( \frac{S_{\rm min}}{\cal S} \right)^{-3/2+\delta} - 16^{-3/2+\delta} & S_{\rm min}/16 < {\cal S} \la S_{\rm min}, \\ 
1 - 16^{-3/2+\delta} & S_{\rm min} \la {\cal S} < S_{\rm max}/16, \\
1- \left( \frac{S_{\rm max}}{\cal S} \right)^{-3/2+\delta}  & S_{\rm max}/16 < {\cal S} < S_{\rm max}. \\
\end{cases} \label{pSnoScint}
\end{eqnarray}
Comparison of eqs.(\ref{SmeasApprox}) and (\ref{pSnoScint}) in the regime $S_{\rm min} \ll {\cal S} \la S_{\rm max}$ reveals the ratio of the scintillation-enhanced to the unenhanced flux density distribution is $\Gamma(5/2-\delta)$.  


\section{Discussion} \label{sec:Compare}
There are two mechanisms by which interstellar scintillation can enhance the detection rate of FRBs at a given flux density, and we discuss the viability of each in turn as an explanation of the disparity in FRB event rates between high and low Galactic latitudes.

The first mechanism is relevant when there is an upper bound to the distribution of FRB flux densities.  The effect of enhancement by a single diffractive scintle is to extend the distribution beyond this cutoff into an exponentially-decreasing tail. Since this tail falls faster than any power law, the mechanism is only of practical relevance if the cutoff is extremely sharp.  We disfavour this mechanism as a practical explanation of the FRB Galactic latitude dependence because it is not clear how the necessary sharp cutoff could arise in practice.  Moreover, the prediction that the differential source counts follows an exponential distribution appears to be at variance with the observed distribution of FRB S/N values.

The second mechanism gives rise to an increase in the detection rate but maintains the intrinsic slope of the differential event rate distribution.  It does not rely on the presence or otherwise of an upper cutoff in the distribution.  Rather, this enhancement comes at the expense of a depressed event rate at the low flux density end of the distribution.  

The phenomenon may be interpreted as an effect of Eddington bias.   The effect of scintillation is to mix populations with different initial flux densities.  Scintillation is equally likely to amplify the radiation as it is to de-amplify it.  However, if the initial flux density distribution is sufficiently steep, the absolute number of low flux density sources redistributed to higher flux densities greatly exceeds the number of high flux density sources redistributed to low flux densities.  When the initial population follows a power law in flux density, the nett effect is to increase the event rate of sources at high flux density relative to those with flux densities near the lower flux density cutoff of the distribution, $S_{\rm min}$.  For flux densities $S_{\rm min} \ll S_\nu \la S_{\rm max}$ the distribution retains the same power-law index as the initial distribution, but the differential event rate is increased by a factor $\Gamma(5/2 - \delta)$.   

\begin{figure}
\centerline{\epsfig{file=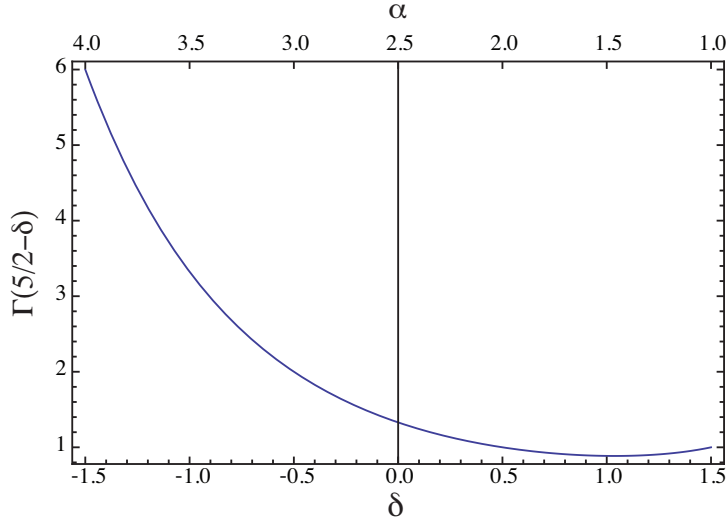, scale=0.7 }}
\caption{The enhancement in the event rate for FRBs subject to diffractive scintillation by a single scintle across the observing band as function of $\delta$ for a flux density distribution scaling as $S_\nu^{-5/2+\delta} \equiv S_\nu^{-\alpha}$.} \label{fig:enhance}
\end{figure}

The nett enhancement is a factor of $\approx 30$\% for a population whose event rate scales as $S_\nu^{-5/2}$ (i.e. for a non-evolving population and neglecting cosmological effects, $\delta =0$).  However, as shown in Figure \ref{fig:enhance}, the enhancement is extremely sensitive to the slope of the distribution: a steeper distribution with $\delta = -1/2$ yields an enhancement of a factor of 2.0, while $\delta = -1$  increases the event rate by a factor of 3.3 over the initial rate.  On the other hand, the effect works in the opposite direction for distributions shallower than $S_\nu^{-2}$, with $\delta = 1/2$ yielding no nett enhancement in event rate, and $\delta = 1$ causing an $\approx 11$\% decrement in the event rate.  

Diffractive interstellar scintillation explains in principle the observed disparity in the detection rate of FRBs at high and low Galactic latitudes.  Accepting this explanation as viable, we can then infer limits on the steepness of the FRB event rate flux density distribution.
Of the 9 FRBs detected to date by the HTRU survey, only 2 have been detected a latitudes $<30^\circ$ and yet the on-sky time at these low latitudes is 23\% higher than at high latitudes. This leads to a 4.7:1 ratio between FRB rates
above and below $30^\circ$ (see also Petroff et al.\,2014). Although 
subject to Poisson statistics, we can place a lower bound on the ratio of high- to low-latitude event rates of 3:1. This in turn implies that FRB event rate distribution must scale more steeply than $S_\nu^{-3.4}$.

The strong departure from the source count index associated with a homogeneously distributed population (whose index is $\alpha=2.5$) suggests that the parent population of FRBs evolves strongly over cosmic time.  Although we do not compare here the implied source count distribution against the predictions of specific models, we remark that such evolution would not be unexpected in a model in which the parent population is tied to the star formation rate.

\subsection{Implications for FRB detection rates}
The model poses an important implication for the FRB progenitor population. 
In particular, the FRB rate derived by Thornton et al. (2013) was based
on the observed rate at high Galactic latitudes. As subsequent observations
have shown (Petroff et al. 2014) the observed rate at low Galactic 
latitudes is about 4 times lower. Our model shows that the rate at low
latitudes is a better reflection of the true rate, and the Thornton rate
should therefore be revised down by a factor of $\sim$4.
At low observing frequencies, the boost in the rate due to scintillation
does not happen as $\Delta \nu_{\rm dc}$ is very much smaller than
at 1.4~GHz and so the revised rate is more appropriate. Coupled with
our uncertain knowledge of the spectral index of FRBs, this may explain
the lack of FRB discoveries so far at low frequencies.

The model also predicts a steep flux density distribution with the
implication that an order of magnitude increase in the event rate at
1.4~GHz could be achieved with a factor 2.6 increase in sensitivity and 
that sensitivity is more important than field-of-view.  The Arecibo
telescope has the highest sensitivity for FRB searches; we estimate
that the detection rate using the Arecibo multibeam at 1.4 GHz should be 
some 14 times that of Parkes surveys according to our model. At high
latitudes Arecibo should detect $\sim$1 FRB per day and should detect
$\sim$1 FRB every 4 days at low latitudes.
Spitler et al. (2014) detected 1 FRB in 12 days observing at low latitudes
with Arecibo, not inconsistent with our prediction especially given
the difficulty of disentangling FRBs from pulsar phenomenon at these
latitudes.

Our model also bodes well for the future of FRB surveys with interferometers
such as the JVLA, ASKAP (Macquart et al.\,2010) and the SKA providing the searches
can be done coherently at high time resolution over the entire field of view.


The progression from the weak- to strong-scintillation limit is expected to follow a trend from high to low Galactic latitudes.  However, we do not present a detailed prediction of the event rate on Galactic latitude. The low number of FRB detections does not yet merit such a detailed comparison.  Moreover, we do not expect there to be a straightforward mapping between enhancement and Galactic latitude.  The turbulent scattering properties of the ISM are known to be highly inhomogeneous and, while there is a general trend to decreasing scattering strength with Galactic latitude, they also depend on Galactic longitude and other details particular to each individual line of sight.  This is particularly pertinent here because there is a strong selection bias to detect only FRBs that are subject to weaker diffractive scintillation; in other words, we expect FRBs to be detected preferentially along sight-lines with anomalously weak scattering. 

%

\section{Conclusion} \label{sec:Conclusions}

Our conclusions are as follows:
\begin{itemize}
\item[--] Galactic diffractive interstellar scintillation can explain the observed disparity in event rates of FRBs between high and low Galactic latitude without altering the slope of the flux density distribution.  The enhancement caused by scintillation is greatest when the decorrelation bandwidth of the scintillation is comparable to or greater than the observing bandwidth.  As the decorrelation bandwidth decreases, the enhancement in the event rate caused by scintillation diminishes.

\item[--] The maximum enhancement in event rate over the intrinsic event rate is $\Gamma(5/2-\delta)$ for an event flux density distribution that scales as $S_\nu^{-5/2+\delta}$.   A consequence of the model is that the frequency of FRB progenitors is a factor $\approx \Gamma(5/2 -\delta)$ lower than supposed based on the observed FRB event rate, which is dominated by the rate count at high Galactic latitudes.  Given the $> $3:1 event rate disparity between high and low-latitude FRBs, the model places a limit of $\delta < -1$ on the steepness of the flux density distribution.   
  
\item[--] The steepness of the event rate distribution implied by the scintillation model implies that only a minor improvement in sensitivity is required to detect FRBs at substantially higher rates.  The integrated event rate for a telescope with a field of view $\Omega$ and sensitive to bursts down to a flux density $S_0$ scales as $\Omega \, S_0^{-3/2+\delta}$.  This strong dependence on $S_0$ argues that sensitivity is a more important criterion for the detection of FRBs than instantaneous field of view. 

\end{itemize}

\section*{Acknowledgments}
Parts of this research were conducted by the Australian Research Council Centre of Excellence for All-sky Astrophysics (CAASTRO), through project number CE110001020.  We are grateful to Steve Spangler whose insightful review improved this manuscript.  We acknowledge stimulating conversations with Ue-Li Pen and Bryan Gaensler.

\end{document}